\documentclass[dvips]{imsart}

\usepackage{amsthm,amsmath}
\usepackage[margin=1.25in]{geometry}

\arxiv{1302.5493}

\startlocaldefs
\usepackage{pdfpages}
\def\bSig\mathbf{\Sigma}

\newcommand{\bA}{\mbox{\boldmath $A$}}
\newcommand{\bG}{\mbox{\boldmath $G$}}
\newcommand{\bX}{\mbox{\boldmath $X$}}
\newcommand{\bY}{\mbox{\boldmath $Y$}}

\newcommand{\bvarepsilon}{\mbox{\boldmath $\varepsilon$}}

\newcommand{\bbeta}{\mbox{\boldmath $\beta$}}

\newcommand{\bv}{\mbox{\boldmath $v$}}
\newcommand{\bI}{\mbox{\boldmath $I$}}
\newcommand{\bS}{\mbox{\boldmath $S$}}
\newcommand{\bZ}{\mbox{\boldmath $Z$}}
\newcommand{\bB}{\mbox{\boldmath $B$}}
\newcommand{\bW}{\mbox{\boldmath $W$}}
\newcommand{\bSigma}{\mbox{\boldmath $\Sigma$}}
\newcommand{\tgamma}{\widetilde{\gamma}}
\newcommand{\hgamma}{\widehat{\gamma}}
\newcommand{\hbbeta}{\widehat{\mbox{\boldmath $\beta$}}}

\endlocaldefs

\begin{document}

\begin{frontmatter}

\title{Modeling and Testing for Joint Association Using a Genetic Random Field Model\thanksref{t1}}
\thankstext{t1}{This paper has been published on Biometrics}
\runtitle{Genetic Random Field Model}


\author{\fnms{Zihuai} \snm{He*}\corref{}\ead[label=e1]{zihuai@umich.edu}},
\author{\fnms{Min} \snm{Zhang**}\corref{}\ead[label=e2]{mzhangst@umich.edu}},
\author{\fnms{Xiaowei} \snm{Zhan}}
\and
\author{\fnms{Qing} \snm{Lu}}
\address{*\printead{e1}}
\address{**\printead{e2}}

\runauthor{Zihuai He et al.}
\affiliation{University of Michigan}

\begin{abstract}
Substantial progress has been made in identifying single genetic variants predisposing to common complex diseases. Nonetheless, the genetic etiology of human diseases remains largely unknown. Human complex diseases are likely influenced by the joint effect of a large number of genetic variants instead of a single variant. The joint analysis of multiple genetic variants considering linkage disequilibrium (LD) and potential interactions can further enhance the discovery process, leading to the identification of new disease-susceptibility genetic variants. Motivated by development in spatial statistics, we propose a new statistical model based on the random field theory, referred to as a genetic random field model (GenRF), for joint association analysis with the consideration of possible gene-gene interactions and LD. Using a pseudo-likelihood approach, a GenRF test for the joint association of multiple genetic variants is developed, which has the  following advantages: 1. accommodating complex interactions for improved performance; 2. natural dimension reduction; 3. boosting power in the presence of LD; 4. computationally efficient. Simulation studies are conducted under various scenarios. The development has been focused on quantitative traits and robustness of the GenRF test to other traits, e.g., binary traits, is also discussed. Compared with a commonly adopted kernel machine approach, SKAT,  as well as other more standard methods,  GenRF shows overall comparable performance and better performance in the presence of complex interactions. The method is further illustrated by an application to the Dallas Heart Study.
\end{abstract}



\end{frontmatter}

\maketitle

\section{Introduction}
\label{s:intro}

With the advance of high-throughput technologies, high-dimensional genetic data have been widely used in  association studies for the identification of genetic variants contributing to common complex diseases. While a large number of genetic variants have been revealed today to be individually associated with complex diseases, they only explain a small proportion of heritability (Manolio, et al., 2009).  Complex diseases are likely influenced by the joint effect of genetic variants through complex biology pathways, given the fact that genes are the functional sets.  However, the multiple testing problem occurs when one considers a set of single locus analyses, which dramatically diminishes  power. Therefore, the joint analysis of a functional set of genetic variants simultaneously can further enhance the discovery process, leading to the identification of new genetic variants associated with complex diseases (Chatterjee, et al. 2006). While the conventional linear or logistic regression models can easily be used for joint association analyses, they are subject to several issues, such as multiple-collinearity, when dealing with a large ensemble of dense genetic markers. The exponentially increased number of parameters also makes them impractical to model two-way or high-order interactions among a large number of genetic variants (Ritchie, et al., 2001).

Several new statistical methods have been recently developed for joint association analysis, including the kernel machine based method (well known as SKAT)(Wu, et al.,  2010;  Wu, et al. 2011) and the similarity regression (SIMreg) (Tzeng, et al. 2009). 
 Both methods significantly reduce the number of  regression parameters, making it feasible and computationally efficient to handle high-dimensional variants. In addition, they  account for linkage disequilibrium (LD) and potential interactions, which further improve performance.  Both SKAT and SIMreg can be thought of as being developed from the general idea that, if genetic association exists, then genetic similarity leads to trait similarity, which is also the intuition behind our method. 


In this paper, we propose a random field framework for modeling and testing for the joint association of multiple genetic variants. We view outcomes as stochastic realizations of a random field on a genetic space and propose to use a random field model, referred  to as a genetic random field model (GenRF), to model the joint association. This approach is motivated by development in spatial statistics where outcomes are viewed as stochastic realizations of a random field on a  Euclidean space (Cressie, 1993). 
This perspective leads to a very distinctive model from the aforementioned methods; specifically, GenRF regresses the response of one subject on responses of all other subjects. GenRF can  be understood from the intuition that genetic similarity   leads to trait similarity if variants are associated with the trait. Under the GenRF model, testing for the joint association reduces to a  test involving a scalar parameter. Using the pseudo-likelihood method, a test for the joint association  is developed, which enjoys many appealing features as SKAT and can achieve comparable or better performance  than existing methods, as demonstrated by simulation studies in Section 3 and a real data application in Section 4. Much of the development is focused  on quantitative traits and robustness of the  test to other traits, e.g., binary traits, is also discussed.

There is a long history of applying  spatial statistical methods to the analysis of genetic data (e.g., Thomas, et al., 2003; Molitor, et al., 2003a; Molitor, et al., 2003b; Iorio and Verzilli, 2007). For example, Iorio and Verzilli (2007) used a spatial probit model to account for the local spatial correlation between variants physically close for fine-scale mapping of disease genes. Molitor, et al. (2003b) used spatial clustering techniques for fine-scale gene mapping. Probably the most closely related work to ours is Molitor, et al. (2003a), which used a spatial auto-regressive model for analysis of haplotypes effects and gene mapping. It differs from our work in two ways. First, it is haplotype-based whereas ours is genotype-based. Second, it is developed from the Bayesian framework where the trait is related to haplotypes through a linear model and the spatial model is used to model the prior distribution of haplotype effects, whereas our method  directly models traits using a spatial model  via a frequentist approach. In this paper, we focus on  multi-marker association testing and the direct spatial modeling of traits using a frequentist approach leads to a test that is analytically tractable and easy to implement.

\section{Method}
\label{s:model}
\subsection{Genetic Random Field Model}
Consider a study where $n$ subjects are sequenced in a region of interest. For subject $i, i=1,\ldots,n$, let $\bG_i$  denote the genotype for the $p$ variants within the region, $Y_i$ the trait or phenotype, and  $\bX_i$ the other covariates  including, for example, demographic and environmental factors.   We are interested in studying the joint association between variants $\bG_{i}$ and trait $Y_{i}$, possibly adjusted for the effect of $\bX_i$.    

As SKAT and SIMreg, our method is also motivated by the general idea that, if the genetic variants are jointly associated with a trait, then the genetic similarity across subjects will contribute to the trait similarity. To put it in another way, if variants are jointly associated with the trait, then the response of a subject would be close to the response of other subjects who share similar genetic and possibly other variables. Based on this key idea, we propose to directly model the response of each subject as a function of all other responses and the contribution of other responses to $Y_i$ is weighted by their genetic  similarity.


For simplicity, we temporarily assume  $Y_i$'s  are centered (have mean zero) and there are no other adjustment covariates.   Specifically, based on the idea discussed above, we model the conditional distribution of $Y_i$ given all other responses as  
\begin{equation}
Y_i  | \bY_{-i} \sim \gamma \sum_{j\neq i}s(\bG_i,\bG_j)Y_{j}+\varepsilon_i,
\label{GenRF}
\end{equation}
where  $\bY_{-i}$ denotes responses for all other subjects except $Y_i$ ; $s(\bG_i,\bG_j)$ is known weights, weighting the contribution of $Y_j$ on  approximating (or predicting) $Y_i$ via their genetic similarity; $\gamma$ is a non-negative coefficient measuring the magnitude of the overall contribution, further discussed below; and $\varepsilon_i$'s are random errors.  A proper weight function $s(\bG_i,\bG_j)$  gives higher value when the two subjects are  more similar  in terms of  genetic variants and, as discussed below,  can be viewed as a measure for proximity of two subjects in a genetic space. The random errors $\varepsilon_i$'s are assumed to be independent and identically distributed with Normal$(0,\zeta^2)$;  extension to distributions other than normal is discussed in Section~\ref{s:GenRFT}.

A main distinction between model (\ref{GenRF}) and the usual regression is that (\ref{GenRF}) models the conditional distribution of $Y_i$ given traits of other subjects, whereas in the usual regression one models the conditional distribution of a subject's  traits  given  his/her genetic variants.  Intuitively,  model (\ref{GenRF}) states that the trait of a subject can be approximated by traits of other subjects who are similar in genetic variants, if variants are associated with the trait. The coefficient $\gamma$ indicates the magnitude of the trait similarity as a result of genetic similarity. Thus, $\gamma$ can also be interpreted as a measure for the magnitude of the joint association of $\bG_i$ with $Y_i$. Specifically, if $\bG_i$ is not associated with $Y_i$, then regardless of how similar subject $i$ is to other subjects in terms of their genetic variants,  the trait $Y_i$ is independent of all other $Y_j$'s for $j\neq i$; that is, $\gamma=0$. On the contrary, if $\bG_i$ is strongly associated with $Y_i$, then one may expect  $Y_i$ can largely be predicted by  traits of subjects having the same or similar genetic variants and a large $\gamma$ indicates a strong joint association.  Therefore, we can test the joint association of genetic variants with the trait by testing a null hypothesis involving a single parameter, i.e., $H_0: \gamma=0$. 

Models like (\ref{GenRF}), where responses are regressed on responses themselves, are referred as auto-regressive models and are commonly used in spatial statistics. In this article, we view the trait as a random field on a genetic space, and from this perspective, model (\ref{GenRF})  is formally a conditional auto-regressive (CAR) model (Cressie, 1993). 
A random field is a generalization of the notation of a stochastic process (Adler and Taylor, 2007).  Informally, a stochastic process is a set of random variables indexed by integers or real numbers. 
A random field can be defined in more general spaces with the index set being an Euclidean space of dimension greater than one or other spaces. For example, in spatial statistics, crop yields of regions can be viewed as  a random field defined in a two-dimensional space.  Regions that are closer in location have  more similar crop yields if spatial correlation exists. For our problem, we may view observed traits as realizations of a random field defined in a $p$-dimensional space of the $p$ genetic variants; that is, corresponding to each ``location'' in the $p$-dimensional genetic space, there is a random response variable associated with it.  Similarly, responses from locations that are ``closer'' in the genetic space are expected to be more similar if the genetic association exists.  In this sense, our model is a  generalization of the  auto-regressive model in  spatial statistics. Models like (\ref{GenRF}) were firstly studied in the seminar work of Besag (1974) for random fields and we will term our model (\ref{GenRF}) as a genetic random field (GenRF) model. As a matter of fact, the GenRF model is closely related to the CAR model in spatial statistics; that is $s(\bG_i,\bG_j)$ analogously defines the proximity of neighbor $\bG_j$ to $\bG_i$ and $\gamma$ is the counterpart of a spatial dependence parameter. However,  we note that the usual tests of spatial dependence, for example, the Cliff-Ord-test (Cliff and Ord, 1972) and the Lagrange Multiplier test (Burridge, 1980),  do not apply in our setting to test for the joint association of variants. The reason is that the matrix $\bS$, defined below, in our GenRF model does not satisfy the regularity condition usually assumed in spatial statistics for deriving the asymptotic distribution, as each subject has infinite neighbors in the genetic space.

We have yet to define a measure for ``closeness'' in the genetic space. Suppose each component of $\bG_i$ records  the number of minor alleles in a single locus and takes on values $\{0, 1, 2\}$, respectively, corresponding to $\{AA, Aa, aa\}$. Then a sensible measure for closeness or similarity is the so called identity-by-state (IBS) (Wu, et al., 2010), defined as 
\begin{eqnarray*}
s(\bG_i, \bG_j)=\sum_{k=1}^p\{2-|G_{ik}-G_{jk}|\}.
\end{eqnarray*}
That is, the IBS measures the number of alleles  in the region of interest shared by two individuals;  for example, for $p=1$,
$s(AA, AA)=2, s(Aa,aa)=1, s(AA,aa)=0$.
 Other measures for closeness in the genetic space rather than IBS are also possible, e.g., the other kernel functions discussed in Wu, at al. (2010),  providing flexibility in our GenRF model. Similar to  SKAT, our GenRF model can also incorporate weights to increase the importance of rare variants. Specifically, one can define $s(\bG_i, \bG_j)=\sum_{k=1}^p w_k\{2-|G_{ik}-G_{jk}|\}$, where $w_k$ is a prespecified weight for variant $k$; see Wu, et al., (2011) for more discussions on $w_k$.

 So far we have focused on the situation where  no covariate adjustment is required. If adjustment for other factors  is needed a natural extension of model (\ref{GenRF}) is given by
\begin{equation}
Y_i  | \bY_{-i}, \bX_i \sim \bbeta^T\bX_i + \gamma \sum_{j\neq i}s(\bG_i,\bG_j)(Y_{j}-\bbeta^T\bX\!_j)+\varepsilon_i.
\label{GenRF2}
\end{equation}
An intercept term is included in $\bX_i$ and, as a result, in (\ref{GenRF2}) $Y_i$'s are not required to be centered. 
Under this model, testing for the joint association of $\bG_i$ with $Y_i$ after adjusting for other factors  is also equivalent to testing $H_0: \gamma=0$. We will mainly focus on  this more general form of the GenRF model in the development of a testing procedure. For simplicity, the matrix form of the GenRF model is given by

\begin{equation}
\bY | \bY_-, \bX = \bX \bbeta+ \gamma \bS(\bY-\bX \bbeta)+\bvarepsilon, 
\end{equation}
where $\bY$ is $(Y_1,\ldots, Y_n)^T$; $\bY_-$ is $(\bY_{-1},\ldots, \bY_{-n})^T$; $\bX$ is an $n\times q$ matrix defined as $(\bX_1^T,\ldots, \bX_n^T)^T$; $\bvarepsilon \sim$ Normal $(0, \zeta^2 I_{n\times n})$; and $\bS$ is an $n\times n$ symmetric matrix with zeros on the diagonal and the $(i,j)$-th element $s(\bG_i,\bG_j)$ for $i\neq j$.


   
According to the factorization theorem of Besag (1974),  our GenRF model in (\ref{GenRF2})   uniquely determines  the following joint distribution of $\bY$,  i.e.,
\begin{eqnarray}
 \bY|\bX \sim \bX\bbeta+\bv, \,\,\, \bv \sim N(0,\zeta^2(\bI-\gamma \bS)^{-1}),
\label{GenRF3}
\end{eqnarray}
where $\bv$ is an $n$-dimensional random column vector.  Note, the coefficient $\gamma$ used for describing the conditional expectation of $Y_i$ given others in model (\ref{GenRF}) actually describes the correlations among $Y_i$'s. 
It is clear that, under the null hypothesis that there is no association between $\bG_i$ and $Y_i$, i.e.,  $\gamma=0$, $Y_i$'s are uncorrelated, but if $\gamma> 0$, GenRF  states that $Y_i$'s are positively correlated as a result of having similar genetic variants associated with the trait. 


\subsection{Genetic Random Field Test}
\label{s:GenRFT}
In this subsection, we focus on developing a test for the null hypothesis $H_0: \gamma=0$ based on model (\ref{GenRF2}).  Model (\ref{GenRF2}) states that, given responses from all other subjects and covariates $\bX_i$,  the conditional distribution of $Y_i$ is normal with mean $\bbeta^T\bX_i + \gamma \sum_{j\neq i}s(\bG_i,\bG_j)(Y_{j}-\bbeta^T\bX_j^T)$  and variance $\zeta^2$. We  construct the pseudo-likelihood according to Besag (1975) as 
\begin{eqnarray*}
L_{pd}=\prod_{i=1}^n \bigg\{\frac{1}{\sqrt{2\pi \zeta^2}}\exp\Big[-\frac{1}{2\zeta^2}\big\{Y_i-\bbeta^T\bX_i -\gamma \sum_{j\neq i}s(\bG_i,\bG_j)(Y_{j}-\bbeta^T\bX_j)\big\}^2\Big]\bigg\},
\end{eqnarray*} 
which is a product of the conditional densities of $Y_i$ across $i$. Also according to Besag (1975), assuming $\bbeta$ is known,  one may estimate $\gamma$ by the maximum pseudo-likelihood method.  The estimator for $\gamma$ can be obtained by minimizing $\sum_{i=1}^n\big\{Y_i-\bbeta^T\bX_i -\gamma \sum_{j\neq i}s(\bG_i,\bG_j)(Y_{j}-\bbeta^T\bX_j\big\}^2$, which in matrix notation is equal to 
\begin{eqnarray*}
\{(\bI-\gamma \bS)(\bY-\bX\bbeta)\}^T(\bI-\gamma \bS)(\bY-\bX\bbeta).
\end{eqnarray*}
 The minimization leads to an estimator for $\gamma$ given by
\begin{equation}
\Rightarrow \tgamma=\frac{(\bY-\bX\bbeta)^T\bS(\bY-\bX\bbeta)}{(\bY-\bX\bbeta)^T\bS^2(\bY-\bX\bbeta)}.
\end{equation}
Intuitively one  expects that a large value of $\tgamma$ would give us evidence to reject the null hypothesis that $\gamma=0$. In practice, $\bbeta$ in unknown. We propose to replace  $\bbeta$ by its least square estimator $\hbbeta$ under the null hypothesis,  i.e., $\hbbeta=(\bX^T\bX)^{-1}\bX^T\bY$ which is unbiased for $\bbeta$. Substitute $\hbbeta$  into the expression for $\tgamma$  and straightforward algebra leads to the final test statistic:
\begin{equation}
\hgamma=\frac{\bY^T\bB\bS\bB\bY}{\bY^T\bB\bS^2\bB\bY},
\label{Test}
\end{equation}
where $\bB=\bI-\bX(\bX^T\bX)^{-1}\bX^T$. 
 Again a large value of $\hgamma$  supports the rejection of the null hypothesis.

We next show how the p-value for testing $\gamma=0$ can be obtained based on the test statistic $\hgamma$; i.e., we would like to calculate the probability of $\hgamma$ greater than the observed value of the statistic under the null hypothesis. 
 Suppose $\eta$ is the observed value of the test statistic $\hgamma$. Since $\bB\bS^2\bB$ is positive-definite, we have 
\begin{eqnarray*}
P_{H_0}\bigg(\frac{\bY^T\bB\bS\bB\bY}{\bY^T\bB\bS^2\bB\bY}>\eta \bigg)=P_{H_0}\bigg((\bB\bY)^T(\bS-\eta \bS^2)\bB\bY>0\bigg)
\end{eqnarray*}
As it is assumed that $\varepsilon_i\sim N(0,\zeta^2)$, i.i.d. across $i$, it follows that $\bB\bY \sim N(0,\zeta^2\bB^2)$ under the null hypothesis. On the other hand, the statistic $\hgamma$ in (\ref{Test}) is ancillary to $\zeta^2$ because  $\zeta^2$ in the numerator and denominator cancels out. Therefore, the above equation becomes
\begin{eqnarray*}
P_{H_0}\bigg((\bB\bY)^T(\bS-\eta \bS^2)\bB\bY>0\bigg)=P\bigg(\bZ^T(\bS-\eta \bS^2)\bZ>0\bigg),  
\end{eqnarray*}
where $\bZ$ is an $n\times 1$ random vector following $N(0, \bB^2)$. Applying standard results on the distribution of quadratic form in normal random variables, we have 
\begin{eqnarray*}
\bZ^T(\bS-\eta \bS^2)\bZ \sim \sum_i^n \lambda_i \Phi_i, 
\end{eqnarray*}
where $\Phi_i$'s are  i.i.d  random variables with $ \chi^2_1$ distribution, and $\{\lambda_i\}$ are the eigenvalues of $\bB(\bS-\eta \bS^2)\bB$. The final p-value can be obtained  by Davies' exact method (1980) for the weighted summation of independent Chi-square variables.


The proposed test has several appealing properties. First, due to the analytical form of the test statistic, the computational burden is well controlled.  Second, as $\hgamma$ in (\ref{Test}) is ancillary to $\zeta^2$,  unlike SKAT,  there is no need to plug in a consistent estimator for $\zeta^2$. Third, the proposed method improves power by exploiting  LD  and allowing for possible complex interactions among variants.   Linkage disequilibrium  can cause correlations between variants, especially when we consider nearby loci. Considering similarity in variants can naturally reduce the degree of freedom. In the  extreme case where components of $\bG_i$ are  ``perfectly correlated'', the similarity argument will consider the whole set as a single variable. In addition, genetic variants involved in the disease pathway are more likely to interact with each other than contribute to risk individually, known as the epistatic variants effect.  Specifying two-way interactions in a set of loci is a challenging high-dimensional problem and the situation gets even worse in modeling higher order interactions. Since  GenRF does not directly model the relationship of $\bG_i$ with $Y_i$, the difficulty of modeling complex interactions are circumvented and the interaction effect is naturally  incorporated through measuring genetic similarity. Finally, as SKAT, the  GenRF test can boost power of testing rare variants by increasing their weights by specifying $w_k$ appropriately for variant $k$. 

\subsection{Robustness to Other Distributions}
The derivation of the GenRF test given above is built on the normal distribution assumption. Asymptotically, the proposed test is robust to distributions other than normal with slight modification.  Consider $P_{H_0}\Big((\bB\bY)^T(\bS-\eta \bS^2)\bB\bY>0\Big)$, where  it is now assumed $\bY$ follows an arbitrary distribution with mean zero and possibly heteroscadastic variances. The random quantity $(\bB\bY)^T(\bS-\eta \bS^2)\bB\bY$ is a  quadratic form in $\bB\bY$ (with mean 0) with matrix $\bA=\bS-\eta\bS^2$. Rotar (1974) proved that under sufficiently weak conditions on matrix $\bA$ and for large $n$, $P_{H_0}\big((\bB\bY)^T(\bS-\eta \bS^2)\bB\bY>0\big)$ is close to  $P_{H_0}(\widetilde{\bZ}^T(\bS-\eta \bS^2)\widetilde{\bZ}>0)$, where $\widetilde{\bZ}$ follows $N(0,\bSigma)$ with $\bSigma$ being the covariance matrix of $\bB\bY$. In addition, Gotze and Tikhomirov (1999) gave an upper bound on $\sup_x \big|P_{H_0}\big((\bB\bY)^T\bA\bB\bY<x\big)-P_{H_0}(\widetilde{\bZ}^T\bA\widetilde{\bZ}<x)\big|$. These properties lead to  the  robustness of the GenRF test, with minor modification,  as long as   $\bB\bY$ has expectation zero under the null hypothesis, which is true since  the least squares estimator $\bX(\bX^T\bX)^{-1}\bX^T\bY$ is  unbiased for the mean of $\bY$ regardless of the distribution of $\bY$.  For example, for binary traits,  $\bSigma=\bB\bW\bB$, where $\bW=\textrm{diag}(\mu_1(1-\mu_1),\ldots,\mu_n(1-\mu_n))$ and $\mu_i=\bbeta^T\bX_i$. Then \begin{eqnarray*}
\widetilde{\bZ}^T(\bS-\eta \bS^2)\widetilde{\bZ} \sim \sum_i^n \widetilde{\lambda}_i \Phi_i, 
\end{eqnarray*}
where $\Phi_i$'s   are  i.i.d  random variables with $ \chi^2_1$ distribution, and $\{\widetilde{\lambda}_i\}$ are the eigenvalues of $\bW^{\frac{1}{2}}\bB(\bS-\eta \bS^2)\bB\bW^{\frac{1}{2}}$. The final p-value can be also obtained  by Davies' exact method (1980).
  We comment that, as the score test in SKAT is of similar quadratic form, one would expect that SKAT may share this property as well.

Therefore,  one can directly use the test statistic in (\ref{Test})  for binary traits or traits that have distributions other than normal and the test, with a minor modification on the null distribution considering heteroscedastic variances, would be asymptotically valid. Note, this test corresponds to a model where the trait mean is related to a linear predictor through an identity link and may seem unnatural for binary traits.  However, we argue that the model is mostly viewed as a mean leading to a sensible test.  We  also note that the commonly used trend test for testing genetic associations in an additive genetic model can  be developed from a linear model for the mean of a binary trait (Laird and Lange, 2011), and a linear model is used  for testing genetic associations for a binary trait in Ballard, et al. (2010) as well.  We note that a possible practical issue for binary traits may arise in practice, i.e.,  the estimated means  $\{\widehat{\mu}_i\}$ may be outside of [0,1]  and consequently $\widehat{\bW}^{\frac{1}{2}}$is not well defined. In this case, a  remedy is to truncate the predictions $\{\widehat{\mu}_i\}$ at 0 or 1.  The practical issue may arise when  covariates have a wide support and a very strong effect and is less of a concern otherwise, for example, when covariates are categorical. Certainly, studying other link functions, e.g., the logit link, to avoid this practical problem is important in the future. The validity of the test, corresponding to an identity link, is  further studied by simulations shown in sections 1 and 2 of the supplemental materials. 
 
\section{Simulation Studies}
\label{s:simulation}

  We report results of several simulations, each based on 1000 Monte Carlo (MC) replicates, to evaluate the performance of the GenRF test, relative to existing methods including SKAT.
 Four sets of simulations are conducted to evaluate 1) type-1 error rates under different minor allele frequencies (MAF) and sample sizes, 2) power for common variant  analysis under different LD,  interaction effect, and proportions of causal SNPs,  3) power under scenarios where the causal  SNPs   include rare variants,  and 4) robustness of the GenRF test to different distributions of the response variable.

In the first set of simulations,  we evaluated type-I error rates using sample size $n=50$, $100$, $200$ and $500$ . Genotypes for $p=20$ loci without LD were simulated,   with MAF for each locus 0.005, 0.01, 0.1, or 0.2. Responses were generated  according to 
\begin{eqnarray*}
Y_i=\varepsilon_i, \textrm { where} \quad \varepsilon_i \sim N (0,1), 
\end{eqnarray*}
so that no genetic variant is associated with the trait.

In the second set of simulations,  we evaluated power under scenarios varying in LD, interaction effects, or the proportions of  causal SNPs, setting  $p=20$ and $n=500$. To simulate LD, the $20$ loci were evenly divided into two  regions. For each region,  the haplotype  allele  was simulated one by one with MAF $0.2$ and   correlation coefficient ($\rho$)  between adjacent pair of alleles equal to 0, 0.2, 0.4, 0.8 respectively for each scenario. Genotypes were then generated by summing up two haplotype vectors. This way, all the loci are positively correlated with others in the  same region.  Responses were generated according to 
\begin{eqnarray*}
Y_i=0.2G_{i,5}+0.2G_{i,15}+\varepsilon_i, \textrm { where} \quad \varepsilon_i \sim N (0,1).  
\end{eqnarray*}
That is, variants 5 and 15, belonging to different LD regions,  are associated with the trait.   

To  generate data with complex interactions, we  set  MAF $0.2$, and the LD parameter $\rho=0.4$.
 Data were generated such that two-way interactions exist  between $K$ ($K=1, 2, 3$ or 4) pairs of alleles, with alleles in each pair  belonging to the two different LD regions as described above. 
 Responses were then generated  according to the following model,

\begin{equation}
Y_i=0.2\sum_{k=1}^{K} G_{i,4+k} G_{i,14+k}+\varepsilon_i, \textrm { where } \quad \varepsilon_i \sim N (0,1). 
\end{equation}
We see that these models contain only interactions but  no main effect of each locus. 

To examine the effect of causal proportion, we   set  MAF $0.2$ and $\rho=0.4$.  For each MC data set, $K$ causal SNPs were randomly selected with $K=1, 2, 3,$ or 4,  each corresponding to 5\%, 10\%, 15\% and 20\% causal SNPs. Responses were then generated according to
\begin{equation}
Y_i=0.15\sum_{k=1}^{K} G_{i,B_k}+ \varepsilon_i, \textrm { where } \quad \varepsilon_i \sim N (0,1),
\end{equation}
where $(G_{B_1},\ldots,G_{B_K})$ are  the selected causal SNPs.

 Simulation set 2 has focused on common variants.  The third set of simulations considered  scenarios involving rare variants and the scenarios vary in proportions of causal variants. We set $p=20$, $n=500$,   and $\rho=0$. 
  The 20 SNPs were divided into two regions,  one with 16 rare variants (MAF 0.008) and one with 4 common variants (MAF 0.1). Note, the proportion of  rare variants is chosen according to the Dallas Heart Study .  Two scenarios were considered where traits were associated with: 1) rare variants only  or 2) both common and rare variants. For each scenario, $K$ rare SNPs were causal with  $K=$  1, 2, 4, 6, 8, 10, 12 or 14, i.e.,  $K\times 6.25\%$ SNPs in the rare region are causal. In the scenario that both rare and common variants are  causal, we set one of the common SNP as causal additionally.
 The effect size $\beta$ was set to be a decreasing function of MAF   with $\beta=0.2\times |\log_{10} \textrm{MAF}|$ as  in Wu et al. (2011).  Responses were generated according to the following model,
\begin{eqnarray*}
Y_i=\beta_1\sum_{k=1}^{K}G_{i,k}+\beta_2G_{i,20}+\varepsilon_i, \textrm { where} \quad \varepsilon_i \sim N (0,1), 
\end{eqnarray*}
where $\beta_1=0.2\times |\log_{10}0.008|, \beta_2=0$ for scenario 1 and $\beta_1=0.2\times |\log_{10}0.008|, \beta_2=0.2\times |\log_{10}0.1|$ for scenario 2.

 We considered one  additional scenario  where the 500 subjects' genotypes were simulated based on data from the Dallas Heart Study. For each MC data set,  we randomly  selected one gene, then we randomly choose $10\%, 20\%,\ldots, 80\%$ causal variants from those rare variants with true MAF less than 1\%. Traits were  simulated by
\begin{eqnarray*}
Y_i=\sum_{k=1}^{K}\beta_{B_k} G_{i,B_k}+\varepsilon_i, \textrm { where} \quad \varepsilon_i \sim N (0,1), 
\end{eqnarray*}
where $(G_{B_1},\ldots,G_{B_K})$ are the selected causal variants and $\beta_{B_k}=0.2\times |\log_{10}\textrm{MAF}_{B_k}|$.

In the fourth set of simulations, we further evaluated the robustness of the GenRF test to distributions other than normal, specifically,  exponential, binary and mixture normal distributions. Details on the simulation setup is described in the  supplemental materials. 

 
In terms of type-I error rates, we only evaluated the proposed GenRF test and SKAT. In both GenRF and SKAT, we adopted the IBS kernel and considered both weighted and unweighted (i.e., $w_k=1$) versions; in the weighted version, Beta (1, 25) weight as in Wu, et al. was used (2011). In addition to SKAT, we compared GenRF test to other more standard methods. For common variant scenarios, we included  the principle component regression test (PCR) (Guaderman et al., 2007); the MinSNP test  (Ballard et al., 2010) and the F-test in linear regression model including only  main effects. For scenarios involving rare variants,  the variable-threshold (VT)  test (Price et al., 2010) was  included.


Table 1 shows results for the first set of simulations with different MAF and sample sizes. The GenRF test achieves the type I error rate close to the nominal level. However, SKAT is conservative in some scenarios due to the estimation of nuisance parameters, especially when the sample size is small. Since the GenRF test is an exact test  without asymptotic approximation  under normal assumption,  the type I error rate is better controlled.

Table 2 shows the power of various methods  under common variant scenarios. The first part shows the effect of LD on power.  When LD  does not exist or is low, e.g., $\rho<0.4$, the three linear regression based tests, PCR, MinSNP and F-test ,  are more powerful as expected because the data were generated exactly from a linear model. Among them, the PCR and MinSNP can  exploit  LD and  have increasing power when   LD  is higher. When  LD  is moderate or high, both the GenRF test and SKAT  have  higher or even substantially higher power than the other tests by borrowing information from other loci. The power of the GenRF test is comparable to that of  SKAT. 

The second part  shows results when there are complex interactions between variants but no main effects. Note the LD structure is  the same as that in part 1  with $\rho=0.4$ in which the five methods have comparable power.  Therefore, the power difference is mainly due to the complex interactions. In these scenarios, the linear regression based methods has low power in detecting the joint association. Both GenRF test and SKAT attain much larger power.  Moreover, the proposed GenRF test has larger power than  SKAT in detecting the joint association effect when complex interactions exist. 

The third part shows results when the causal proportion varies. Similarly, the LD parameter $\rho$ is set to be 0.4 to eliminate the impact of factors other than the causal proportion. Because MinSNP is based on single SNP analysis, the test is less powerful especially when causal proportion is high, i.e. $15\%$ or $20\%$. GenRF and SKAT show comparable power in general, but GenRF performs better as causal proportion gets higher.

Table 3 shows results for scenarios involving rare variants . When the trait is only associated with rare variants, the weighted GenRF and SKAT have significantly larger power as we expected because the weights favor the rare variants. The weighted GenRF has lower power than SKAT when the causal proportion is low, e.g., $\le 25\%$, but has larger power then the proportion is greater than $25\%$. Both weighted GenRF and SKAT have comparable or larger power relative to the VT test and F-test. The scenario based on the Dallas Heart Study shows similar results,  i.e. GenRF performs better under higher causal proportion ($\ge 20\%$).

When causal variants include both common and rare variants and the effect size is a decreasing function of MAF, the unweighted GenRF and SKAT have comparably larger power than the weighted tests when the rare causal proportion is low   ($\le 37.5\%$). This is not surprising as the  effect of the common variant is relatively large but  down-weighted in the  weighted GenRF and SKAT. As the rare causal proportion increases and the number of  common variants is fixed at one, the results change  dramatically. When the rare causal proportion is higher than $37.5\%$, the weighted GenRF and SKAT show higher power than the unweighted counterpart.  Overall, for scenarios considered here, the GenRF test has very good performance relative to others.



Supplementary Tables 1 and 2 show the robustness of the  GenRF test to distributions other than normal. In implementing GenRF  and SKAT, the identity link is used in modeling the responses.   The GenRF test achieves the type I error rate close to the nominal level even when the distribution of the response is not normal; the same holds for SKAT. Particularly, for binary traits, we evaluated the robustness of the GenRF test to heteroscedastic variances in Section 2 of the supplementary materials. Valid type I error rates and reasonably good power are achieved. In addition, the remedy by truncation when predictions fall outside of the range of [0,1] works well in practice, even under extreme and possibly unrealistic scenarios.

\begin{table}
\caption{Type I error rate simulation results under different levels of  MAF and sample size (1000 replicates).   Each cell contains the type I error rate, i.e., rejection rate when data are generated under the null model. GenRF: the unweighted genetic random field test; SKAT: the unweighted sequential kernel association test of Wu, et al. (2011); GenRF.w: GenRF with Beta(1,25) weight as in Wu, et al. (2011); SKAT.w:  SKAT with Beta(1,25)  weight.}
\begin{center}
 \small
\begin{tabular}{ccccccccccc}
\hline 
\hline
Methods && \multicolumn{9}{c}{Different Levels of MAF and Sample Size (n) }\\
\hline
 &MAF& \multicolumn{4}{c}{0.005} & & \multicolumn{4}{c}{0.01}\\
 &n & 50 & 100 & 200 & 500 & & 50 & 100 & 200 & 500\\
 \cline{3-6}  \cline{8-11} \\
GenRF & & 0.043 & 0.049 & 0.050 & 0.045 && 0.040 & 0.043 & 0.061 & 0.048\\
GenRF.w & & 0.048 & 0.056 & 0.051 & 0.045 & & 0.043 & 0.046 & 0.060 & 0.046\\\\
SKAT  & &0.035 & 0.051 & 0.057 & 0.057 & & 0.034 & 0.046 & 0.050 & 0.039\\
SKAT.w & & 0.034 & 0.050 & 0.053 & 0.059 & & 0.029 & 0.042 & 0.046 & 0.035\\\\
\hline 
 &MAF  & \multicolumn{4}{c}{0.1} & & \multicolumn{4}{c}{0.2}\\
 &n  & 50 & 100 & 200 & 500 & & 50 & 100 & 200 & 500\\
 \cline{3-6}  \cline{8-11} \\
GenRF & & 0.051 & 0.049 & 0.055 & 0.044 & & 0.050 & 0.058 & 0.055 & 0.049\\
GenRF.w  & & 0.052 & 0.052 & 0.053 & 0.048 & & 0.047 & 0.051 & 0.052 & 0.046\\\\
SKAT & & 0.022 & 0.039 & 0.041 & 0.041 & & 0.016 & 0.030 & 0.041 & 0.043\\
SKAT.w & & 0.041 & 0.035 & 0.043 & 0.054 & & 0.045 & 0.043 & 0.046 & 0.041\\\\
\hline 
\end{tabular}
\label{t:table1}
\end{center}
\end{table}

\begin{table}
\caption{Power simulation results for common variant  analysis  under different levels of linkage disequilibrium (LD), interaction effects and causal proportion (1000 replicates).   GenRF: the unweighted genetic random field test; SKAT: the unweighted sequential kernel association test of Wu, et al. (2011); PCR: the princeple component regression test of Guaderman et al. (2007); MinSNP: the MinSNP test considered by Ballard et al. (2010); F-test: the F-test in linear regression.  }
\begin{center}
 \small
\begin{tabular}{ccccccccccccccc}
\hline 
\hline
Method & \multicolumn{4}{c}{Different Level of LD} &  & \multicolumn{4}{c}{Number of Two-way Interactions} &  & \multicolumn{4}{c}{Different Causal Proportion}\\
\hline 
 & 0 & 0.2 & 0.4 & 0.8 &  & 1 & 2 & 3 & 4 & & 5\% & 10\% & 15\% & 20\% \\
\cline{2-5} \cline{7-10} \cline{12-15} \\
GenRF & 0.462 & 0.472 & 0.566 & 0.816 &  & 0.119 & 0.364 & 0.652 & 0.862 & & 0.124 & 0.321 & 0.539 & 0.776 \\
SKAT & 0.491 & 0.487 & 0.545 & 0.764 &  & 0.100 & 0.299 & 0.546 & 0.746 & & 0.150 & 0.324 & 0.506 & 0.727\\\\
\cline{2-5} \cline{7-10} \cline{12-15} \\
PCR & 0.495 & 0.467 & 0.518 & 0.676 &  & 0.119 & 0.268 & 0.470 & 0.657 & & 0.159 & 0.308 & 0.473 & 0.679\\
MinSNP & 0.570 & 0.507& 0.543 & 0.656 &  & 0.098 & 0.252 & 0.408 & 0.576 & & 0.180 & 0.342 & 0.463 & 0.624\\
F-test & 0.545 & 0.514 & 0.524 & 0.538 &  & 0.112 & 0.231 & 0.394 & 0.562 & & 0.145 & 0.278 & 0.471 & 0.665\\\\
\hline 
\end{tabular}
\label{t:table2}
\end{center}
\end{table}

\begin{table}
\caption{Power simulation results  under scenarios involving rare variants with different proportion of causal variants (1000 replicates). Rare Causal Variants:   causal variants are rare only;  Common $\&$ Rare Causal Variants: causal variants are both rare and common; DHS: scenario based on the Dallas Heart Study.   GenRF.w: the genetic random field test with Beta (1, 25) weight as in  Wu, et al. (2011); SKAT.w: the sequential kernel association test with Beta (1, 25)  weight; VT: the variable-threshold test of Price et al. (2010); Other entries as in Table 2.  }
\begin{center}
 \small
\begin{tabular}{ccccccccc}
\hline
\hline
Method & \multicolumn{8}{c}{Different Proportion of Causal Variants}\\
\hline 
 & \multicolumn{8}{c}{ Rare Causal Variants}\\\\
 & 6.25\% & 12.5\% & 25\% & 37.5\% & 50\% & 62.5\% & 75\% & 87.5\%\\
\cline{2-9} \\
GenRF & 0.045 & 0.073 & 0.139 & 0.209 & 0.305 & 0.466 & 0.593 & 0.739\\
SKAT & 0.048 & 0.052 & 0.066 & 0.073 & 0.086 & 0.100 & 0.111 & 0.126\\\\
GenRF.w & 0.062 & 0.087 & 0.212 & 0.429 & 0.660 & 0.848 & 0.950 & 0.980\\
SKAT.w & 0.083 & 0.125 & 0.252 & 0.368 & 0.515 & 0.654 & 0.736 & 0.814\\\\
VT & 0.065 & 0.082 & 0.128 & 0.209 & 0.314 & 0.487 & 0.680 & 0.852\\
F-test & 0.080 & 0.113 & 0.190 & 0.302 & 0.449 & 0.556 & 0.670 & 0.765\\\\
\hline
\hline
 & \multicolumn{8}{c}{ Common \& Rare Causal Variants}\\\\
 & 6.25\% & 12.5\% & 25\% & 37.5\% & 50\% & 62.5\% & 75\% & 87.5\%\\
\cline{2-9} \\
GenRF & 0.191 & 0.259 & 0.380 & 0.501 & 0.625 & 0.761 & 0.861 & 0.927\\
SKAT & 0.274 & 0.281 & 0.287 & 0.313 & 0.331 & 0.359 & 0.387 & 0.416\\\\
GenRF.w & 0.061 & 0.097 & 0.232 & 0.434 & 0.646 & 0.853 & 0.939 & 0.981\\
SKAT.w & 0.078 & 0.155 & 0.277 & 0.386 & 0.523 & 0.631 & 0.732 & 0.818\\\\
VT & 0.217 & 0.306 & 0.418 & 0.504 & 0.603 & 0.720 & 0.845 & 0.930\\
F-test & 0.163 & 0.270 & 0.354 & 0.477 & 0.618 & 0.701 & 0.779 & 0.843\\\\
\hline 
 & \multicolumn{8}{c}{ DHS}\\\\
 & 10\% & 20\% & 30\% & 40\% & 50\% & 60\% & 70\% & 80\%\\
\cline{2-9} \\
GenRF & 0.080 & 0.140 & 0.169 & 0.247 & 0.329 & 0.414 & 0.507 & 0.600\\
SKAT & 0.071 & 0.089 & 0.114 & 0.117 & 0.153 & 0.191 & 0.205 & 0.271\\\\
GenRF.w & 0.100 & 0.204 & 0.321 & 0.434 & 0.588 & 0.696 & 0.796 & 0.875\\
SKAT.w & 0.118 & 0.196 & 0.294 & 0.330 & 0.433 & 0.544 & 0.600 & 0.688\\\\
VT & 0.095 & 0.159 & 0.254 & 0.359 & 0.498 & 0.612 & 0.721 & 0.827\\
F-test & 0.147 & 0.239 & 0.355 & 0.423 & 0.528 & 0.653 & 0.721 & 0.795\\\\
\hline
\end{tabular}
\label{t:table2}
\end{center}
\end{table}

\section{Application}
\label{s:application}

We  applied our method to the Dallas Heart Study (Browning et al., 2004), a population-based, multi-ethnic study on 3551 subjects whose Lipids and glucose metabolism were measured. In this study, 348 sequence variations in the coding regions of the four genes, ANGPTL3, ANGPTL4, ANGPTL5 and ANGPTL6 were discovered. Most of these variants (86\%) are rare with  MAF  less than 1\%. More information regarding the number of rare variants is  shown in the Supplementary Materials. Individuals who have diabetes mellitus, alcohol dependency or have taken lipids lowering drugs were excluded as these factors may confound the interpretation of associations.  Our final analysis was based on data on 2812 subjects after quality control steps.

We  assessed  the association between ANGPTL gene families and two traits, specifically high-density lipoprotein (HDL) and triglyceride,  using the proposed GenRF test and SKAT, both with and without weighting.  As in the simulation studies, the IBS kernel and the Beta (1, 25) weight were applied. Analyses were also carried out using the more traditional methods including PCR, MinSNP, VT and F-test.
 Our analysis were done for the non-synonymous variants, adjusted for gender and ethnicity.


The association between ANGPTL4 gene and the level of HDL and triglyceride was previously discovered by Romeo, et al. (2007). In our analysis, both weighted GenRF and SKAT gave evidence for the ANGPTL4 and triglyceride association (p-values: 0.019 and 0.006). Among all the methods considered,  only weighted SKAT showed marginal evidence for the association between ANGPTL4 and HDL (p-value: 0.040). One possible explanation is that the causal proportion of ANGPTL4 is low and SKAT performs better in this case as shown in simulation studies. Note that the weighted GenRF and SKAT uncovered these associations while the unweighted tests did not, possibly indicating the causal variants in ANGPTL4 might be rare (MAF $<$ 5\%), or the effect size is negatively correlated  with allele frequency. As for ANGPTL5, our analysis using GenRF provided evidence to support the association with HDL (p-value: 0.009 and 0.036 for weighted and unweighted analyses) while SKAT provided marginal evidence (p-value: 0.035 and 0.050). Note the unweighted tests gave larger p-values. Since all variants in ANGPTL5 are rare (MAF $<$ 5\%), the result suggests that the causal variants might be the rare variants with relatively higher allele frequency. This finding was supported by standard approaches like  MinSNP (p-value: 0.033), F-test (p-value: 0.051) and  VT test (p-value: 0.051).  More results are  shown in table 4. 
Overall, for this  study,  GenRF   performs comparably to SKAT and seems to perform better than the other more standard methods.

\begin{table}
\caption{Application to Dallas Heart Study for non-synonymous variants. GenRF: the unweighted genetic random field test; SKAT: the unweighted sequential kernel association test of Wu, et al. (2011); PCR: the princeple component regression test of Guaderman et al. (2007); MinSNP: the MinSNP test considered by Ballard et al. (2010); F-test: the F-test in linear regression; GenRF.w: the genetic random field test with Beta (1, 25) weight of Wu, et al. (2011); SKAT.w: the sequential kernel association test with Beta (1, 25) weight; VT: the variable-threshold test of Price et al. (2010). $\ast$ indicates p-value is less than or equal to $\alpha=0.05$.}
\begin{center}
 \small
\begin{tabular}{cccccccccccc}
\hline
\hline
Method&&\multicolumn{4}{c} {P-value} \\
 \hline\\
 &&\multicolumn{4}{c} {HDL } \\\\
 & & ANGPTL3 & ANGPTL4 & ANGPTL5 & ANGPTL6\\
 \hline
 GenRF & &0.487&0.181&0.009$\ast$&0.417\\
 SKAT & &0.981&0.423&0.035$\ast$&0.504\\\\
 PCR & &0.980&0.775&0.197&0.434\\
 MinSNP & &0.178&0.329&0.033$\ast$&0.729\\
 F-test & &0.331&0.148&0.051&0.786\\\\
 GenRF.w & &0.345&0.218&0.036$\ast$&0.496\\
 SKAT.w &&0.965&0.040$\ast$&0.050$\ast$&0.535\\
 VT & &0.393&0.111&0.051&0.488\\\\
 
 &&\multicolumn{4}{c} {Triglyceride } \\\\
 & & ANGPTL3 & ANGPTL4 & ANGPTL5 & ANGPTL6\\
 \hline
 GenRF & &0.025$\ast$&0.221&0.428&0.857\\
 SKAT & &0.050$\ast$&0.312&0.936&0.755\\\\
 PCR & &0.129&0.780&0.787&0.762\\
 MinSNP & &0.562&0.219&0.921&0.713\\
 F-test & &0.587&0.380&0.904&0.530\\\\
 GenRF.w & &0.100&0.019$\ast$&0.180&0.466\\
 SKAT.w & &0.075&0.006$\ast$&0.906&0.756\\
 VT & &0.993&0.905&0.968&0.050$\ast$\\\\

\hline
\end{tabular}
\label{t:table3}
\end{center}
\end{table}


\section{Discussion}
\label{s:discuss}

 We have proposed a novel framework for modeling and testing for the joint association of genetic variants with a trait from the perspective of viewing traits as a random field on a genetic space.
The development has been focused on quantitative traits with a normal distribution. 
  Based on the GenRF model, a test for genetic associations was developed and this test enjoys many appealing features. The GenRF test is based on testing a null hypothesis involving a single parameter, allowing it to exploit LD to improve power. When  LD  is moderate or high, our simulations showed that  the GenRF test achieves much higher power than  the more traditional regression-based methods. The GenRF model is flexible to allow for complex interaction effects and,  as demonstrated by simulations,  the  GenRF test is even much more powerful than  SKAT in the presence of complex interaction effects. Moreover, as SKAT, prespecified variant-specific weights can be incorporated  to boost power for rare variants. Unlike SKAT, the GenRF test is an exact test under the normal assumption and thus not overly conservative in finite samples. Finally, the  test is  computationally easy to implement  since an  analytical form is available. In summary,  the GenRF test is an appealing alternative to SKAT and other existing methods for testing the joint association of variants with a trait. It can achieve overall comparable performance  and sometimes even much better performance   relative to SKAT as well as other methods.  
  
Although we focus on quantitative traits, we note that the GenRF test is  robust to distributions other than normal as discussed previously and 
demonstrated  by simulation studies.   Specifically for binary traits, although the GenRF model with an identity link function  may seem a bit unnatural, the  resulting  test with a minor modification is still valid  and  can achieve good power. However, due to  the conceptual difficulty  associated with modeling binary traits using a linear model and the possible practical issue that can arise, it would be interesting to study, within the framework of random field model,  other link functions for binary traits as well as other  distributions in the future.

\section{Supplementary Materials}

\setcounter{equation}{0}

 \renewcommand\tablename{Supplementary Table}

 \newcommand{\logit}{\mbox{\rm logit}}
 \newcommand{\expit}{\mbox{\rm expit}}

 \newcommand{\sqrtn}{n^\frac{1}{2}}
 \newcommand{\isqrtn}{n^{-\frac{1}{2}}}
 \newcommand{\op}{o_p(1)}
 \newcommand{\invn}{n^{-1}}
 \newcommand{\convp}{\stackrel{p}{ \longrightarrow}}

\subsection{Robustness to other distributions}
We evaluated the robustness of the GenRF test to distributions other than normal. The GenRF test for traits with distributions other than normal is described in Section 2.3 of the main manuscript. The simulation setup is otherwise similar to the first set of simulations, described in Section 3 of the main manuscript,  with only one region, $p=10$, $\rho=0.4$ and $n=100$. Responses  $Y_i$ were generated according to generalized linear models using the canonical link function, i.e., 
\begin{eqnarray*}
g(\mu_i)=aG_{i,5},
\end{eqnarray*}
where $a$ was set to be 1.1 and 2.5 respectively for exponential and binary distributions. For Mixture Normal, we generated two normal distributions with  mean difference 10, equal mixture proportions, and $a= 2.7$. We set $a$ to be 0 in evaluating  the type-I error rate.   The results are shown in Supplementary Table \ref{t:webt1}. 


\begin{table}[htbp!]
\caption{Simulation results under different distributions of the response variable (1000 replicates). $\ast$ indicates results are unavailable due to ``sample size is small, need small sample adjustment'' and SKAT has no small sample adjustment for IBS kernel.  }
\begin{center}
 \small
\begin{tabular}{cccccc}
\hline
\hline
 Method&&\multicolumn{3}{c} {Distribution} \\
 \hline
 & & Exponential & Mixture Normal & Binary\\
 \hline
 GenRF & Power&0.636&0.582&0.646\\
  & Type I&0.052&0.056&0.046\\
  SKAT & Power&0.655&0.582&$\ast$\\
  & Type I &0.046&0.046&$\ast$\\
 F-test & Power&0.572&0.568&0.559\\
 & Type I&0.056&0.054&0.050 \\
\hline
\end{tabular}
\label{t:webt1}
\end{center}
\end{table}

\subsection{Robustness to heteroscedastic variances of binary traits}
We evaluated the robustness of the GenRF test to heteroscedasitc variances of binary traits. Since the variance of a binary outcome is a function of its mean, the variance is known to be heteroscedastic when the mean of outcome depends on covariates. The modification of the GenRF test for binary traits is described in Section 2.3 of the main manuscript. The simulation setup is otherwise similar to the first set of simulations, described in Section 3 of the main manuscript, with $p=20$, $\rho=0$, minor allele frequency $0.2$, and $n=100$. Responses  $Y_i$ were generated according to logistic models, i.e., 
\begin{eqnarray*}
\logit(p_i)=aG_{i,5}+bX_i,
\end{eqnarray*}
where $a$ was set to be 3 in evaluating power and 0 in evaluating type I error rate; $X_i$ was a covariate generated from $N(0,1)$; and $b$ was varying from 0 to 10 to generate different levels of heteroscedastic variance. A larger coefficient $b$ results in a wider range of the predicted mean and thus more heteroscedastic variance. When $b=5$ or $10$, which represents unusually strong effect of $X$ (probably unlikely in practice), some predicted means fall outside of $[0,1]$ and truncation at 0 or 1 was used. The results are shown in Supplementary Table \ref{t:webt2}. We note that the power decreases as the coefficient $b$ increases because the noise becomes larger. The type I error is well controlled even if some predicted means reached 0 or 1, indicating that the GenRF test with the minor modification is robust to heteroscedastic variances of binary traits.


\begin{table}[htbp!]
\caption{Simulation results under different levels of heteroscedastic variances (500 replicates). Coefficient: the coefficient of the covariate.  $\ast$ indicates that some predicted means reached 0 or 1.}
\begin{center}
 \small
\begin{tabular}{cccccccc}
\hline
\hline
 Method&&\multicolumn{5}{c} {Coefficient} \\
 \hline
 & & 0 & 1 & 3 & 5$\ast$ & 10$\ast$\\
 \hline
 GenRF & Power&0.538&0.594&0.408 &0.290 &0.110\\
  & Type I&0.052&0.046&0.040 &0.058 &0.060\\

\hline
\end{tabular}
\label{t:webt2}
\end{center}
\end{table}

\subsection{Application to Dallas Heart Study}
We analyzed data from the Dallas Heart Study (Browning et al., 2004.), a population-based, multi-ethnic study on 3551 subjects whose lipids and glucose metabolism are measured. In this study, sequence variations in the coding regions of the four genes, ANGPTL3, ANGPTL4, ANGPTL5 and ANGPTL6 are discovered. Supplementary Table \ref{t:webt3}  lists the number of non-synonymous variants in each gene and their MAFs.

 \begin{table}[htbp!]
  \renewcommand\arraystretch{0.7}
 \caption{Dallas Heart Study sequencing data information: number of non-synonymous variants in each gene. MAF: minor allele frequency.}
 \begin{center}
  \small
 \begin{tabular}{cccccccccccc}
 \hline
 \hline
 &&\multicolumn{4}{c} {Number of Variants } \\
  \hline\\
  & & ANGPTL3 & ANGPTL4 & ANGPTL5 & ANGPTL6\\
  \hline
  All  & &21&25&18&25\\
  MAF $<$ 5\% & &21&24&18&25\\
  MAF $<$ 1\% & &20&23&17&24\\
 \hline
 \end{tabular}
 \end{center}
 \label{t:webt3}
 \end{table}
 
 \section*{Acknowledgements}
The authors would like to thank Jonathan Cohen for the permission to use the Dallas Heart Study data and Dajiang Liu for preparing the data. The authors also highly appreciate Michael Boehnke's comprehensive suggestions, and thank for the valuable comments from Xihong Lin, Seunggeun Lee, William Wen, Veronica Berrocal, Laura Scott, Lu Wang, Dajiang Liu, Bhramar Mukherjee and Hui Jiang.

\end{document}